\documentstyle[11pt]{article}

\renewcommand{\theequation}{\arabic{section}.\arabic{equation}}

\begin{document}

\font\ninerm = cmr9

\baselineskip 14pt plus .5pt minus .5pt

\def\footnoterule{\kern-3pt \hrule width \hsize \kern2.6pt}

\hsize=6.0truein
\vsize=9.0truein
\textheight 8.5truein
\textwidth 5.5truein
\voffset=-.4in
\hoffset=-.4in

\pagestyle{empty}
\begin{center}
{\large\bf String-Inspired Gravity Coupled 
to Yang-Mills Fields}\footnote{\ninerm
\hsize=6.0truein
This work was
supported in part by funds provided by 
the European Community under contract \#ERBCHBGCT940685,
the U.S. 
Department of Energy (D.O.E.)
under contract \#DE-AC02-89ER40509,
the Korea Science and Engineering Foundation through
the SRC program of SNU-CTP,
the Basic Science Research Research Program
under project \#BSRI-96-2425,
and the Istituto Nazionale di Fisica Nucleare (INFN).}
\end{center}

\vskip 1.5cm

\begin{center}
{\bf G. Amelino-Camelia$^{(a)}$, D. Bak$^{(b)}$, 
and D. Seminara$^{(c)}$}\\
\end{center}

\begin{center}
{\it (a)  Theoretical Physics, University of Oxford,
1 Keble Rd., Oxford OX1 3NP, UK} \\
{\it (b)  Department of Physics,
Seoul City University,
Seoul 130-743, Korea} \\
{\it (c) Center for Theoretical Physics,
Laboratory for Nuclear Science, and Department of Physics,
Massachusetts Institute of Technology,
Cambridge, Massachusetts 02139, USA}
\end{center}

\baselineskip=12pt
\vspace{1.2cm}
\begin{center}
{\bf ABSTRACT}
\end{center}
String-inspired 1+1-dimensional 
gravity is coupled to
Yang-Mills fields
in the Cangemi-Jackiw
gauge-theoretical
formulation, based on the extended 
Poincar\'e  group.
A family of couplings,
which involves metrics obtainable from the physical metric
with a conformal rescaling, is considered,
and the resulting family of models is investigated
both at the classical and the quantum level.
In particular, 
also using a series of Kirillov-Kostant phases,
the wave functionals that solve the constraints
are identified. 

\bigskip
\bigskip
\bigskip
\begin{center}
Physical Review D54 (1996) 6193
\end{center}
\bigskip
\bigskip
\vfill
MIT-CTP-2539 \space\space
OUTP-96-18-P \space\space
SNUTP-96-044 \space\space\space\space
hep-th/9611028
\hfill
March 1996
\eject

\pagestyle{plain}
\pagenumbering{arabic}
\setcounter{page}{1}

\pagebreak[3]
\setcounter{equation}{0}
\renewcommand{\theequation}{\arabic{section}.\arabic{equation}}
\section{Introduction}
\nopagebreak
\medskip
\nopagebreak
\indent

The problem of constructing a consistent quantum theory of gravity
has proven to be extremely hard. It appears that, 
once there is a quantum dynamics for the geometry,
very few of the tools 
used in the quantization of 
theories in a background geometry are available.

One appealing possibility to tackle the problem is the one of
casting gravity in a gauge theoretical formulation,
so that we would be able to draw from the experience
gained in the many successful quantizations of gauge theories.
Some of the most interesting proposals providing such formulations
are
the Ashtekar formulation of Einstein gravity \cite{Asht}, the
Poincar\'e gravities \cite{Ver1,Gri1}, 
the Chern-Simons gravity \cite{Witt},
and, most recently, the Cangemi-Jackiw \cite{Can92} 
gauge theoretical
reformulation, based on the extended 
Poincar\'e  group, of 
string-inspired (1+1-dimensional) gravity \cite{CGHS}.

The Cangemi-Jackiw approach to 1+1-dimensional 
quantum gravity
has been used in investigations
of pure gravity \cite{Can92},
gravity coupled to point particles \cite{Can93,bs}, 
and gravity coupled to
scalar matter fields\cite{Can94}; however, the analysis of 
gravity coupled to
gauge fields, which is the objective of the present paper,
had not been previously performed.
Obviously, for gauge 
theoretical formulations of quantum gravity the coupling
to gauge fields can be very interesting; most importantly,
one expects simplifications
(with respect to corresponding non gauge theoretical 
formulations\cite{kun,strobl,kungauge})
to arise, allowing to make substantial progress.

One important aspect of our analysis is that we 
consider different ways to couple gauge fields to gravity.
We consider a family of couplings
involving metrics
that can be obtained from the Cangemi-Jackiw gauge metric
with a  conformal rescaling,
so, in particular, we have as limiting cases the 
minimal coupling
via the gauge metric itself and
the minimal coupling via the physical
metric, as done in Ref. \cite{bs}. 
(The definitions of the gauge metric and the physical
metric are reviewed in the following section.)
In particular,
the investigation of the coupling via the physical
metric might be relevant to the understanding of
the nature of the divergencies
encountered in Ref. \cite{bs} in relation to the 
Poincar\'e coordinates.
A crucial point is that in the case of gauge-theoretical
gravity coupled to $N$ point particles \cite{bs}
one is naturally lead to the introduction of $N$ sets
of Poincar\'e coordinates
associated to the actual coordinates of the particles
(upon appropriate gauge choice the Poincar\'e coordinates
are indeed the coordinates of the particles),
whereas the coupling to fields always involves one set of
Poincar\'e coordinates,
which however are then functions taking values on the
entire 1+1-dimensional space-time.

\noindent
Throughout the paper, 
the abelian limit of the results that we obtain
for arbitrary Yang-Mills fields
within the gauge-theoretical formulation are compared
to the corresponding results obtained
within the geometric approach of Ref.\cite{kungauge},
in which only the coupling to
abelian gauge fields was considered.

Before proceeding to the quantization
of our models, which is the primary objective of this paper,
for completness
in the next two sections we 
review the Cangemi-Jackiw gauge theoretical formulation 
and analyze our models at the classical level.

\pagebreak[3]
\setcounter{equation}{0}
\renewcommand{\theequation}{\arabic{section}.\arabic{equation}}
\section{Gauge formulation of lineal gravity}
\nopagebreak
\medskip
\nopagebreak
\indent

The (geometrical) action
of string-inspired gravity \cite{CGHS}
is given by
\begin{equation}
\label{GG}
I=\frac{1}{2\pi \kappa}\int d^2 x \sqrt{- g^{\!_P}} e^{-2\phi} 
( R(g^{\!_P})+
4 g^{\!_P \mu\nu}\partial_\mu \phi\partial_\nu \phi -\lambda).
\end{equation}
where $\lambda$ is the cosmological constant and $\phi$  the 
dilaton field.\footnote{\hsize=6.0truein 
Notation: the signature of  the metric 
tensor $g^{\!_P}_{\mu\nu}$ is assumed 
to be $(1,-1)$. The Latin indices ${a,~b,~c\dots}$ run over a 
tangent space 
where the flat Minkowski metric $h_{ab}={\rm diag}(1,-1)$ is defined.
The antisymmetric 
symbol $\epsilon^{ab}$ is normalized so that $\epsilon^{01}=1$.} 
The 
introduction of the new variables 
\begin{equation}
\label{eq2}
g_{\mu\nu}=e^{-2\phi}  g^{\!_P}_{\mu\nu} 
\ \ \ \ {\rm and}\ \ \ \ \eta=e^{-2\phi}
\end{equation}
transforms the action (\ref{GG}) into a simpler expression
\begin{equation}
\label{FF}
I=\frac{1}{2\pi \kappa}\int d^2 x \sqrt{-g} (\eta R(g) -\lambda),
\end{equation}
which can be reformulated as a gauge theory \cite{Ver1,Can92,Gri94}. 
(In the following, in order to avoid ambiguities, we shall refer to
$g^{\!_P}_{\mu\nu}$ as the {\it ``physical''} metric, and to $g_{\mu\nu}$
as the {\it ``gauge''}  metric.)
In particular, a gauge theoretical formulation of the action (\ref{FF})  
can be given  by using the 4-parameter extended Poincar\'e 
group in 1+1 dimensions \cite{Can92,Can93,Can94}, whose Lie algebra
reads
\begin{eqnarray}
\label{algebra}
&&[P_a,P_b]=\epsilon_{ab} I\,,\ \ \ \ \ \ \ [P_a,J]=\epsilon_{a}^{~b} 
P_b \\
\end{eqnarray}
Here, $P_a$ and $J$ are the usual translation and boost generators,
while $I$ is the central element.
Such  extension arises naturally in two dimensions
if one  allows non-minimal gravitational coupling, as pointed 
out in Ref.\cite{Can92}. 

\noindent
The  field, which will describe gravity, is now introduced as a connection
one-form that takes values in the Lie algebra 
\begin{equation}
\label{connection}
B_\mu=e^a_\mu P_a +\omega_\mu J+a_\mu I.
\end{equation}
$e^a$ and $\omega$ are the {\it zweibein} and the 
spin connection respectively; the potential $a_\mu$ is, instead, 
related to the volume form \cite{Can92}. The
connection defined in (\ref{connection}) transforms according to 
the adjoint representation. In components the transformation is
\begin{eqnarray}
&&e^a_\mu \rightarrow (\Lambda^{-1})^a_{~b}(e^b_\mu+\epsilon^b_{~c}
\theta^c\omega_\mu
+\partial_\mu\theta^b),\\
&&\omega_\mu \rightarrow \omega_\mu +\partial_\mu \alpha,\\
&&a_\mu \rightarrow  a_\mu-\theta^a\epsilon_{ab} e^b_\mu-
\frac{1}{2}\theta^a\theta_a \omega_\mu+\partial_\mu\beta
+\frac{1}{2}\partial_\mu\theta^a\epsilon_{ab}\theta^b,
\end{eqnarray}
where we have parameterized the gauge transformation as follows
\begin{equation}
U={\rm exp}(\theta^a P_a) ~ {\rm exp}(\alpha J) ~ {\rm exp}(\beta I)
\label{gaugep}
\end{equation}
and $\Lambda^a_{~b}$ is the Lorentz transformation matrix
\begin{equation}
\Lambda^a_{~b}=\delta^a_{~b} \cosh\alpha+\epsilon^a_{~b}\sinh\alpha.
\end{equation}
\noindent
The field strength ${\cal R}$ can be  now computed from its definition
\begin{eqnarray}
\label{curvature}
&&{\cal R}=dB+[B,B]\nonumber\\
&&=(de^a +\epsilon^a_{~b}\omega\wedge e^b) P_a+ d\omega J+ (da+
\frac{1}{2}\epsilon_{ab} e^a\wedge e^b) I.
\end{eqnarray}
To construct an invariant action 
linear in the curvature (i.e. in ${\cal R}$),  
one introduces a multiplet, $\eta_A\equiv (\eta_a,\eta_2,\eta_3)$,
that transforms according to the co-adjoint representation
\begin{eqnarray}
&&\eta_a \rightarrow (\eta_b
-\eta_3\epsilon_{bc}\theta^c)\Lambda^b_{~a},\\
&&\eta_2 \rightarrow \eta_2-\eta_a\epsilon^a_{~b}\theta^b+
\frac{1}{2}\eta_3 \theta^a\theta_a,\\
&&\eta_3 \rightarrow \eta_3\,.
\end{eqnarray}
(Note that $\eta_a$ may be set to zero by a gauge transformation.)
The  action is now simply formed by contracting $\eta_A$ with 
$\epsilon^{\mu\nu} {\cal R}^A_{\mu\nu}$
\begin{equation}
\label{pio}
I_g=\frac{1}{2\pi \kappa}
\int d^2 x \epsilon^{\mu\nu}\Bigl(\eta_a 
(\partial_\mu e^a_\nu + \omega_\mu \epsilon^a_{~b} e^b_\nu ) +
\eta_2 \partial_\mu\omega_\nu+\eta_3 (\partial_\mu a_\nu+\frac{1}{2}
\epsilon_{ab} e^a_\mu e^b_\nu)\Bigr)\,.
\end{equation}
It is easy to show \cite{Can92,Can93} that this {\it B-F} theory is 
equivalent to the string-inspired gravity defined by the geometrical action 
(\ref{FF}) once  we identify $\eta_2=2 \eta $ and $ g_{\mu\nu}=
e^a_\mu e_{a\nu}$. The cosmological constant, $\lambda$, is generated 
dynamically by the field $\eta_3$, which is fixed to be a constant by 
the equations of motion.  

\noindent
A gauge invariant description of matter requires the introduction of 
a new variable: the Poincar\'e coordinate $q^a$. The appearance of
this additional  degree of freedom is intrinsically  related to 
the geometric
structure of the Poincar\`e group. However, a detailed analysis of 
this subject goes beyond the aim of this brief review, and for 
a deeper analysis we refer the reader to Refs.\cite{Stelle,Dresh,Kawai}. 
Here, we only comment on some aspects useful in
writing down invariant actions for matter fields.

\noindent 
In a Poincar\'e gauge theory of gravity, the connection $e^a_\mu$
cannot be really interpreted as the geometrical {\it zweibein}.
In fact, due
to the inhomogeneous nature of its transformation under the symmetry,
the ensuing metric ($g_{\mu\nu}(x)=e^a_\mu\eta_{ab}e^b_\nu$) would be
not gauge invariant\footnote{This unpleasant feature may disappear
when the equations of motion impose the vanishing of the curvature
fields. In such cases the symmetry may be recovered on-shell.  This
happens, for example, for the pure gravity described by the action 
(\ref{pio}).}. Hence, to preserve a geometrical interpretation,  we
need  to construct a new field that plays the role of the
geometrical {\it zweibein}.

\noindent
Assuming that $q^a$ under a gauge transformation behaves like
\begin{equation}
\label{pp}
(q^U)^a=(\Lambda^{-1})^a_{~b}(x)(q^b(x) +
\epsilon^b_{~c}\theta^c(x)),
\end{equation}
the combination $E^a_\mu(q) \! \equiv \! 
- \epsilon^{a}_{~b} 
\Bigl[\partial_\mu q^b(x)+\epsilon^b_{~c}
\Bigl(q^c \omega_\mu-e^c_\mu \Bigr) \Bigr]$
seems to be a good canditate.  In fact it has the correct
transformation law, (namely $E^a_\mu(q)=\Lambda^a_b(\alpha) 
E^b_\mu(q)$). Moreover, there is a gauge choice in which 
$E^a_\mu(q)$ can be actually identified with $e^a_\mu$,
the so-called {\it ``physical gauge''} $q^a=0$.  When this gauge 
is selected, the Poincar\`e invariance is broken and only the Lorentz 
subgroup survives, so that the usual first fomalism is recovered.

\noindent
In a certain sense the $q^a$ field looks like a Higgs field in a gauge
theory with symmetry breaking; its presence insures the gauge
invariance, and when the unitary gauge, $q^a \! = \! 0$,
is chosen the physical content
of the theory is exposed.

\noindent
The construction of gauge invariant actions for matter fields can be
now performed in a straightforward manner. Given the geometrical
Lagrangian in the first order formalism, we shall simply replace the
field $e^a_\mu $ with the field $E^a_\mu (q)$ everywhere it appears.
This procedure will naturally lead to the desired Lagrangian.

\noindent
For instance, we can consider the case of a massless scalar field,
whose action in the usual first order formalism can be written as
\begin{equation}
\int_{\cal M} dx^2\epsilon_{ab}
\epsilon^{\mu\nu} (e^a_\mu \Pi^b\partial_\nu \phi-
e^a_\mu \phi\partial_\nu \Pi^b+ e^a_\mu e^b_\nu \Pi_l \Pi^l).
\end{equation}
The field $\Pi^l$ is an auxiliary field introduced in order to 
have a polynomial Lagrangian  in  the {\it zweibein} $e^a_\mu$.  
The Poincar\`e invariant Lagrangian is now simply
\begin{equation}
\int_{\cal M} dx^2\epsilon_{ab}
\epsilon^{\mu\nu} (E_\mu^a (q) \Pi^b\partial_\nu \phi-
E_\mu^a (q) \phi\partial_\nu \Pi^b+ E_\mu^a(q) E_\nu^b(q)\Pi_l \Pi^l).
\end{equation}

\noindent
Finally, concerning the equations of motion derived from such
Lagrangians, we have to notice that  they are in general consistent 
only when considered together with the equations for the  gravity.
This feature is common to all Poincar\`e theories of gravity.

\pagebreak[3]
\setcounter{equation}{0}
\renewcommand{\theequation}{\arabic{section}.\arabic{equation}}
\section{The model and its classical solutions}
\nopagebreak
\medskip
\nopagebreak
\indent

In this section we shall investigate the string-inspired  gravity
coupled to  a non abelian gauge field $A_\mu(x)=A^i_\mu(x)T_i$. In 
the geometric formulation, where the metric, the dilaton and the 
gauge connections are the fundamental fields, the action reads
\begin{eqnarray}
\label{eq1}
I&=&\frac{1}{2\pi \kappa}\int d^2 x \sqrt{-g_{\! _P}} 
(S^{\! _P}(\phi) \, R(g^{\! _P})+ 
g^{\! _P \mu\nu} \, \partial_\mu \phi\partial_\nu \phi 
+ V^{\! _P}(\phi))
\nonumber\\
&-&\frac{1}{4}\int d^2 x 
\sqrt{- g^{\! _P}} \, W^{\! _P}(\phi) \, {\rm Tr}(F_{\mu\nu} F^{\mu\nu}).
\end{eqnarray}
where $R(g^{\! _P})$ is the scalar curvature 
and $F_{\mu\nu}=\partial_\mu A_\nu
-\partial_\nu A_\mu +[A_\mu,A_\nu]$. With respect to the usual dilaton 
gravity, we have added the possibility of an arbitrary dilaton potential 
$V^{\! _P}(\phi)$ and arbitrary couplings to the
dilaton $(S^{\! _P}(\phi) \ {\rm and }\  W^{\! _P}(\phi))$.
If $S^{\! _P}(\phi)$ is a regular
invertible function for any  admissible value of 
$\phi$, the geometrical action can be connected  to  the gauge formulation 
in terms of the extended Poincar\`e group by means of the following field 
redefinition
\begin{equation}
g_{\mu\nu}(x)= {\rm exp} \left ( \frac{1}{2} \int
\frac{d\phi}{d S^{\! _P}/d\phi} \right )
g^{\! _P}_{\mu\nu}(x)\ \ \ \ \ \
\bar \phi = S^{\! _P}(\phi),
\end{equation}
In terms of this new fields, the action takes the form 
\begin{equation}
\label{eq4}
I=\frac{1}{2\pi \kappa}\int d^2 x \sqrt{- g} (\bar\phi R(g) -
\lambda) + \frac{1}{8} \int d^2 x 
\sqrt{-g}\left( W(\bar\phi) {\rm Tr}({\tilde F}^2) -2 V(\bar\phi)\right )
\end{equation}
where ${\tilde F} \equiv \epsilon^{\mu\nu}F_{\mu\nu}$, and
$V$ and $W$ are defined as
\begin{eqnarray}
V(\bar\phi) &=& \frac{2\lambda}{\pi \kappa}
- \frac{2}{\pi \kappa}
V^{\! _P}(\phi (\bar\phi)) \,
{\rm exp} \left ( - \frac{1}{2} \int
\frac{d\phi}{d S^{\! _P}/d\phi} \right ) \\
W(\bar\phi) &=& { W^{\! _P}(\phi (\bar\phi))} \,
{\rm exp} \left ( \frac{1}{2} \int
\frac{d\phi}{d S^{\! _P}/d\phi} \right )
\end{eqnarray}
The connection with the gauge theory is now rather simple; in fact,
we have
\begin{eqnarray}
\label{action2}
&&I_g=\frac{1}{2\pi \kappa}\int d^2 x  \epsilon^{\mu\nu} \Bigl(
\eta_a (\partial_\mu e^a_\nu + \omega_\mu \epsilon^a_{~b} e^b_\nu ) +
\eta_2 \partial_\mu\omega_\nu+\eta_3 (\partial_\mu a_\nu+\frac{1}{2}
\epsilon_{ab} e^a_\mu e^b_\nu)\Bigr) \\
&&-\frac{1}{4}\int dx^2 \Biggl \{ V(q^A\eta_A){\rm det}(E(q)) -
W(\eta_A q^A) \left ( {\rm Tr}({\tilde F} \Phi)
-\frac{{\rm det}(E(q))}{2} {\rm Tr }
(\Phi \Phi)\right ) \Biggr\}\nonumber
\end{eqnarray}
where $q^A\eta_A $ is the gauge invariant 
combination $\eta_a q^a +\eta_2 +\frac{1}{2} \eta_3 q_a q^a$, 
and ${\rm det}(E(q))
=-1/2\epsilon_{ab}\epsilon^{\mu\nu}E^a_\mu(q) E^b_\nu(q)$.
The auxiliary field $\Phi=\Phi^i T_i$ has been introduced in order to 
have a polynomial Lagrangian.   
The equivalence between the two actions
(\ref{eq4}) and (\ref{action2}) can be easily shown by comparing 
the equations of motion in the two theories. In the gauge-theoretical
formulation the equations of motion read 
\begin{eqnarray}
\label{etaa}
&&\!\!\!\!\!\!\!\!
\delta\eta_a \rightarrow \epsilon^{\mu\nu}
(\partial_\mu e^a_\nu + \omega_\mu \epsilon^a_{~b} e^b_\nu ) 
= \frac{\pi \kappa}{2} q^a 
\biggl [\frac{\partial  V}{\partial(\eta^A q_A)}{\rm det}(E(q))   
- \frac{\partial W}{\partial(\eta^A q_A)}  
{\cal T}\biggr ],\\
\label{eta2}
&&\!\!\!\!\!\!\!\!
\delta\eta_2\rightarrow 
\epsilon^{\mu\nu}\partial_\mu \omega_{\nu}
= \frac{\pi \kappa}{2}\left [ 
\frac{\partial  V}{\partial(\eta^A q_A)} 
{\rm det}(E(q)) - \frac{\partial W}{\partial(\eta^A q_A)}  
{\cal T}\right],\\
\label{eta3}
&&\!\!\!\!\!\!\!\!
\delta\eta_3\rightarrow  \epsilon^{\mu\nu} 
(\partial_\mu a_\nu +\frac{1}{2} 
\epsilon_{ab} e^a_\mu e^b_\nu)=\frac{\pi \kappa}{2} q^l q_l\left [  
\frac{\partial  V}{\partial(\eta^A q_A)} 
{\rm det}(E(q)) - \frac{\partial W}{\partial(\eta^A q_A)}  
 {\cal T} \right ], ~~~~~~~~~~~~~ \ \  \ \ \ \ \\
\label{ea}
&&\!\!\!\!\!\!\!\!
\delta e^a\rightarrow \partial_\mu 
\eta_a+\epsilon_{ab} \eta^b\omega_\mu
+\epsilon_{ab} e^b_\mu\eta_3=-\frac{\pi \kappa}{2} \left (
  V(\eta_A q^A) + \frac{ W (\eta_A q^A)}{2}{\rm Tr }(\Phi \Phi)
\right) \epsilon_{ab}E^b_\mu(q),\ \ \ \ \\
\label{omega}
&&\!\!\!\!\!\!\!\!
\delta\omega_\mu \rightarrow \partial_\mu 
\eta_2+\epsilon_{ab}\eta^a e^b_\mu
=-\frac{\pi \kappa}{2} \left (
 V(\eta_A q^A) + \frac{ W (\eta_A q^A)}{2}{\rm Tr }(\Phi \Phi)
\right) \epsilon_{ab}E^a_\mu (q) q^b,\\
\label{a}
&&\!\!\!\!\!\!\!\!
\delta a_\mu\rightarrow \partial_\mu \eta_3=0,\\ 
\label{phi}
&&\!\!\!\!\!\!\!\!
\delta \Phi  \rightarrow {\tilde F} -
{\rm det}(E(q))  \Phi=0 ,\\
\label{Amu}
&&\!\!\!\!\!\!\!\!
\delta A_\mu \rightarrow  {\cal D}_{\mu} (\Phi  W (\eta_A q^A))=0,  
\end{eqnarray}
where ${\cal D}_{\mu}$ is the covariant 
derivative constructed with the Yang-Mills field $A_\mu$,
and ${\cal T}$ stands for 
\begin{equation}
{\cal T}=\left ( {\rm Tr}({\tilde F} \Phi)
-\frac{1}{2}{\rm det}(E(q)){\rm Tr }(\Phi \Phi)\right ).
\end{equation}
The auxiliary field $\Phi$ can be now eliminated by using (\ref{phi}).
In fact we have
\begin{equation}
\label{gac1}
\Phi= \frac{{\tilde F}}{{\rm det}(E(q))},  
\end{equation}
which in turn implies
\begin{equation}
\label{gac2}
{\rm Tr}(\Phi\Phi)=\frac{{\rm Tr}({\tilde F}^2)}{{\rm det}
(E^a(q))^2}\ \ \ \  {\rm and}\ \ \ \ {\cal T}=
\frac{1}{2}\frac{{\rm Tr}({\tilde F}^2)
}{{\rm det}(E(q))} 
~.
\end{equation}
Moreover, from Eq.(\ref{Amu}) it is straightforward to show 
that the quantity
\begin{equation}
\label{gac3}
Q \equiv
\frac{{\rm Tr}({\tilde F}^2)}{{\rm det}(E(q))^2 } W(\eta^A q_A)^2 \ \ 
\end{equation}
is constant ($x$-independent).
Using (\ref{gac1})-(\ref{gac3}), 
the equations for gravity become
\begin{eqnarray}
\label{etaa1}
&&\epsilon^{\mu\nu} (\partial_\mu e^a_\nu
+ \omega_\mu \epsilon^a_{~b} e^b_\nu ) = \frac{\pi \kappa}{2} q^a 
\biggl [\frac{\partial  V}{\partial(\eta^A q_A)}
+ \frac{\partial W}{\partial(\eta^A q_A)}  
\frac{Q}{2 W(\eta^A q_A)^2}
\biggr ]{\rm det}(E(q)),\ \ \ \ \ \ \ \ \\
\label{eta21}
&&\epsilon^{\mu\nu}\partial_\mu \omega_{\nu}
= \frac{\pi \kappa}{2}\left [ 
\frac{\partial  V}{\partial(\eta^A q_A)} 
+\frac{\partial  W}{\partial(\eta^A q_A)}  
\frac{Q}{2 W(\eta^A q_A)^2}\right]{\rm det}(E(q)),\ \ \ \ \ \ \ \\
\label{eta31}
&&\epsilon^{\mu\nu} (\partial_\mu a_\nu +\frac{1}{2} 
\epsilon_{ab} e^a_\mu e^b_\nu)=\frac{\pi \kappa}{2} q^l q_l\left [  
\frac{\partial  V}{\partial(\eta^A q_A)} 
+\frac{\partial  W}{\partial(\eta^A q_A)}  
\frac{Q}{2  W(\eta^A q_A)^2} \right ] {\rm det}(E(q)),\ \ \ \ \
\ \ \ \ \\
\label{ea1}
&&\partial_\mu \eta_a+\epsilon_{ab} \eta^b\omega_\mu
+\epsilon_{ab} e^b_\mu\eta_3=-\frac{\pi \kappa}{2} \left (
  V(\eta_A q^A) + \frac{Q}{2 W(\eta^A q_A) }
\right)\epsilon_{ab}E^b_\mu(q),\\
\label{omega1}
&&\partial_\mu \eta_2+\epsilon_{ab}\eta^a e^b_\mu
=-\frac{\pi \kappa}{2} \left (
 V(\eta_A q^A) + \frac{Q}{2 W (\eta_A q^A)}
\right)\epsilon_{ab} E^a_\mu (q) q^b,\\
\label{a1}
&&\partial_\mu \eta_3=0.
\end{eqnarray} 
Let us introduce an effective dilaton potential defined by
\begin{equation}
\hat V_Q(\eta^A q_A)= V(\eta_A q^A) + \frac{Q}{2 W (\eta_A q^A)} \,.
\end{equation}
With this choice, the previous set of equations takes a 
simpler form
\begin{eqnarray}
\label{etaa2}
&&\epsilon^{\mu\nu} (\partial_\mu e^a_\nu 
+ \omega_\mu \epsilon^a_{~b} e^b_\nu ) = \frac{\pi \kappa}{2} q^a 
\frac{\partial \hat V}{\partial(\eta^A q_A)}{\rm det}(E(q)),\\
\label{eta22}
&&\epsilon^{\mu\nu}\partial_\mu \omega_{\nu}= \frac{\pi \kappa}{2}
\frac{\partial \hat V}{\partial(\eta^A q_A)} {\rm det}(E(q)),\\
\label{eta32}
&&\epsilon^{\mu\nu} (\partial_\mu a_\nu +\frac{1}{2} 
\epsilon_{ab} e^a_\mu e^b_\nu)=\frac{\pi \kappa}{2} q^l q_l  
\frac{\partial \hat V}{\partial(\eta^A q_A)} 
{\rm det}(E(q)),\\
\label{ea2}
&&\partial_\mu \eta_a+\epsilon_{ab} \eta^b\omega_\mu
+\epsilon_{ab} e^b_\mu\eta_3=-\frac{\pi \kappa}{2} 
 \hat V_Q(\eta_A q^A)
\epsilon_{ab}E^b_\mu(q),\\
\label{omega2}
&&\partial_\mu \eta_2+\epsilon_{ab}\eta^a e^b_\mu
=-\frac{\pi \kappa}{2}
\hat V_Q(\eta_A q^A) 
\epsilon_{ab}E^a_\mu (q) q^b,\\
\label{a2}
&&\partial_\mu \eta_3=0,
\end{eqnarray} 
which are the equations of motion for gravity in absence
of the gauge field. Therefore the  effect of $A^i_\mu$ is
only to change the shape of the dilaton potential
from $ V(\eta^A q_A)$ to  $\hat V_Q(\eta^A q_A)$.

\noindent
Combining now Eq.(\ref{etaa2}) with Eq.(\ref{eta22}), we
can easily derive that
\begin{equation}
\label{tau}
\partial_\mu E^a_\nu(q)-\partial_\nu E^a_\mu (q)
+ \omega_\mu \epsilon^a_{~b} E^b_\nu(q)
- \omega_\nu \epsilon^a_{~b} E^b_\mu(q)
=0 ~,
\end{equation}
namely the geometrical {\it zweibein} $E^a_\mu(q)$ is torsionless.
Notice that this property, which is fundamental if we want 
to have only a metric theory of gravity, does not hold 
for the gauge connection $e^a_\mu$ (see Eq.(\ref{etaa2})).
(Actually $e^a_\mu$ becomes torsionless only in the 
physical gauge $q^a=0$, see Eq.(\ref{etaa2}) again.)
This confirms what we stated in the previous section.

\noindent  From now on, we focus our attention on the classical 
solutions of the reduced system (\ref{etaa2})--(\ref{a2}). First of all 
we  note  that Eq.(\ref{a2})  requires $\eta_3$ to be a constant,
and we call its value ``$\lambda$'' to get agreement with the 
geometric description in Eq.(\ref{eq4}).  Moreover, we can also 
neglect Eq.(\ref{eta32}) because it simply fixes  the potential 
$a_\mu$, which does not play a role in the discussion of the 
geometry.

\noindent
Combining Eqs.(\ref{ea2})-(\ref{a2}),  it is straightforward
to show that
\begin{equation}
\label{3.32}
\partial_\mu(\eta_A  q^A)=-
\lambda\epsilon_{ab}\left (q^a+\frac{\eta^a}{\lambda}\right ) E^b_\mu.
\end{equation}
and 
$$
\partial_\mu (\eta_A\eta^A)\!=\!-{\pi \kappa}\lambda\hat V_Q(\eta_A q^A)
\epsilon_{ab}\left (q^a+\frac{\eta^a}{\lambda}\right
) E^b_\mu\!=\!{\pi \kappa}\hat V_Q(\eta_A q^A)
\partial_\mu(\eta_A q^A)\!=\!
\partial_\mu\left( {\pi \kappa}J_Q(\eta_A q^A)\right)
$$
\begin{equation}
\label{3.33}
\Longrightarrow \ \ \ \eta_A\eta^A= \xi +{\pi \kappa}J_Q(\eta_A q^A)
\end{equation} 
where $J_Q(x)$ is a $x$-primitive of the function $\hat V_Q(x)$,
$\xi$ is a constant, and $\eta_A \eta^A$ is the gauge
invariant combination $\eta_a \eta^a-2\eta_2\eta_3$.

\noindent
Taking the covariant derivative $\nabla_\nu$ of Eq.(\ref{3.32}),
we get, with the help of Eq.(\ref{ea2}),
\begin{equation}
\label{dfdf}
\nabla_\nu\partial_\mu (\eta_A q^A)=-\lambda g_{\mu\nu}(q)+
\frac{\pi\kappa}{2} \hat V_Q(\eta_A q^A) g_{\mu\nu}(q).
\end{equation}
Note that, importantly, in this equation
both the metric $g_{\mu\nu}(q)$ and the covariant derivative
$\nabla_\mu$ are constructed out of the geometrical {\it zweibein}
$E^a_\mu(q)$.

\noindent
Upon selecting the conformal gauge $E^a_\mu(q)=\delta^a_\mu
e^{\sigma(x)}$\footnote{Notice that this gauge is always available,
due to the diffeomorphism and Lorentz invariance}, 
Eq.(\ref{dfdf}) takes the form
\begin{equation}
\partial_\mu\partial_\nu (\eta_A q^A)-
\partial_\nu (\eta_A q^A)\partial_\mu \sigma -
\partial_\mu (\eta_A q^A)\partial_\nu \sigma -
\eta_{\mu\nu}\partial^\lambda (\eta_A q^A)\partial_\lambda \sigma =
[ \frac{\pi\kappa}{2} \hat V_Q(\eta_A q^A) -\lambda ]
e^{2\sigma}\eta_{\mu\nu}.
\end{equation}
The component $++$ and $--$ ($x^+=x+t$ and $x-=x-t$) of this equation 
can be, now, casted in the following way
\begin{equation}
\partial_+( e^{-2\sigma}\partial_+ (\eta_A q^A))=0
\ \ \ \ \ \ \ \ 
\partial_-( e^{-2\sigma}\partial_- (\eta_A q^A))=0,
\end{equation}
which are very easy to solve 
\begin{equation}
\partial_- (\eta_A q^A)= e^{2\sigma} f(x^+) ~, \ \ \ \ \ \
\partial_+ (\eta_A q^A)= e^{2\sigma} g(x^-) ~.
\label{gacstar}
\end{equation}
Here, $f(x^+)$ and $g(x^-)$ are two arbitrary functions, which
can be set to $1$ by using the residual difffeomorphism invariance
present in the conformal gauge ($x^+\to x^+(\tilde x^+)$
and $x^-\to x^-(\tilde x^-)$). With this choice 
the Eqs.(\ref{gacstar})
collapse to
\begin{equation}
\label{ccc1}
\frac{\partial (\eta_A q^A)}{\partial t}= 0 ~, \ \ \ \ \ \
\frac{\partial (\eta_A q^A)}{\partial x}= e^{2 \sigma} ~.
\end{equation}
Now, with the help of Eq.(\ref{3.32}), Eq.(\ref{3.33}) can be 
rewritten as follows
\begin{equation}
g^{\mu\nu}\partial_{\mu} (\eta_A q^A)\partial_{\nu} (\eta_A q^A)+
2 \lambda (\eta_A q^A)+{\pi \kappa} J_Q(\eta_A q^A)+\xi=0.
\end{equation}
In conformal gauge this equation, due to Eq.(\ref{ccc1}), reduces to 
\begin{equation}
\label{ccc2}
\frac{\partial(\eta_A q^A)}{\partial x}+2 \lambda (\eta_A q^A)+
{\pi \kappa}J_Q(\eta_A q^A)+ \xi =0,
\end{equation}
which can be integrated easily with respect to $\eta_A q^A$, giving
\begin{equation}
x=-\int^{\eta_A q^A} dy \frac{1}{2\lambda y+{\pi \kappa} J_Q(y)+ \xi }.
\end{equation}
This equation fixes implicitely $\eta_A q^A$ in terms of the
coordinate $x$. In turn, from Eq.(\ref{ccc1}),
we can compute the conformal factor $\sigma(x)$. Given these  two
quantities the geometry is completely determined.

\noindent
It is easy to show that no further constraint arises from the
remaining equations. In fact Eq.(\ref{eta22}) is implied
by  the Eqs.(\ref{ccc1}) and (\ref{ccc2}), once the 
condition of vanishing torsion (\ref{tau}) is taken into account.
Finally, Eqs.(\ref{ea2}) and (\ref{omega2}) simply determine
$e^a_\mu$ and $\eta_2$, if we fix $\eta_a=0$  by using the 
invariance under Poincar\`e translations.
The Poincar\`e coordinate $q^a$ is, instead, determined by solving 
the equation $E^a_\mu(q) \! \equiv \! - \epsilon^{a}_{~b} 
\Bigl[\partial_\mu q^b(x)+\epsilon^b_{~c}
\Bigl(q^c \omega_\mu-e^c_\mu \Bigr) \Bigr] \! = \! e^\sigma \delta^a_\mu$.

\noindent  
The last step is the construction of the gauge field. Combining 
the equations of motion with the gauge invariance, we can always 
choose a solution of the form
\begin{equation}
F^i_{\mu\nu}=\frac{f^i}{2 W(\eta_A q^A)} {\rm det}(E(q))
\epsilon_{\mu\nu}     
\end{equation}
where $f^i$ is a constant vector with the property $f^i f_i = Q$.

\noindent
As a closing remark on the analysis at the classical level, 
we notice that our results are consistent with (and generalize to the
nonabelian case)
those of Ref.\cite{kungauge}, where coupling of dilaton
gravity to a $U(1)$ gauge field was
investigated in the framework of the geometrical formulation. 

\pagebreak[3]
\setcounter{equation}{0}
\renewcommand{\theequation}{\arabic{section}.\arabic{equation}}
\section{Quantization}
\nopagebreak
\medskip
\nopagebreak
\indent

We now turn to the quantization of our model. 
We begin by recording the Lagrange density that is the 
starting point of the analysis
\begin{eqnarray}
\label{actionq1}
&&{\cal L} = \frac{1}{2\pi \kappa}  \epsilon^{\mu\nu} \Bigl(
\eta_a (\partial_\mu e^a_\nu + \omega_\mu \epsilon^a_{~b} e^b_\nu ) +
\eta_2 \partial_\mu\omega_\nu+\eta_3 (\partial_\mu a_\nu+\frac{1}{2}
\epsilon_{ab} e^a_\mu e^b_\nu)\Bigr) \\
&&-\frac{1}{4} 
 V(q^A\eta_A){\rm det}(E(q)) 
+ \frac{1}{4} W(\eta_A q^A) 
\left ( {\rm Tr}({\tilde F} \Phi)
-\frac{{\rm det}(E(q))}{2} {\rm Tr }
(\Phi \Phi)\right ) 
\nonumber
\end{eqnarray}
This can be rewritten in a way more clearly exposing the symplectic 
structure as follows
\begin{eqnarray}
{\cal L} \!\!\!&=&\!\!\! {1 \over 2 \pi \kappa}
\left( \eta_a \dot{e}_1^a
+ \eta_2 \dot{\omega}_1
+ \eta_3 \dot{a}_1 \right)
+ \frac{1}{4} \left( V(\eta_A q^A) + W(\eta_A q^A){ 
\frac{1}{2} {\rm Tr} (\Phi\Phi)} \right)
E_{a1} \dot{q}^a 
\nonumber\\
& &
+ e_0^a G_a + \omega_0 G_2 + a_0  G_3    
+ \frac{1}{4}{\rm Tr}( W(\eta_A q^A)\Phi \dot{A}_1)
- \frac{1}{4}{\rm Tr}({A}_0 {\cal D}_1 ( W(\eta_A q^A)\Phi))
\label{primadipa}
\end{eqnarray}
where spatial, but not temporal, integration  by parts has been
carried out freely, and $G_a$, $G_2$, $G_3$ are the gravitational 
gauge generators
\begin{eqnarray}
G_a &\equiv& {1\over 2 \pi \kappa}
\left(
\eta'_a + \epsilon_a^{~b} \eta_b \omega_1 + \eta_3 \epsilon_{ab} e_1^b
\right)
+ \epsilon_a^{~b} p_b
\label{eq12a} \\
G_2 &\equiv& {1\over 2 \pi \kappa}
\left( \eta'_2 + \eta_a \epsilon^a_{~b} e_1^b \right)
- q^a \epsilon_a^{~b} p_b
\label{eq12b} \\
G_3 &\equiv& {1\over 2 \pi \kappa} \eta'_3
~.\label{eq12c}
\end{eqnarray}
The symbol $p_a$ in Eqs.(\ref{eq12a}), (\ref{eq12b}) and (\ref{eq12c})  
stands for the expression 
\begin{eqnarray}
\frac{1}{4}\left( V(\eta_A q^A) + \frac{1}{2} W(\eta_A q^A){ 
{\rm Tr}(\Phi\Phi)}\right) E_{a1}  ,
\end{eqnarray}
which is the coefficient of $\dot{q}^a$ in the Lagrangian.
It is obviously convenient to promote $p_a$
to a momentum conjugate to $q_a$, by 
introducing
a Lagrange multiplier $u^a$ that enforces the explicit form of $p_a$
Analogously, it is convenient
to promote the coefficient of $\dot A_1$ to
a momentum $\Pi$ conjugate to $A_1$.
\begin{eqnarray}
{\cal L} &=& {1 \over 2 \pi \kappa}
\left( \eta_a \dot{e}_1^a
+ \eta_2 \dot{\omega}_1
+ \eta_3 \dot{a}_1 \right)
+ p_a
\dot{q}^a 
+ e_0^a G_a + \omega_0 G_2 + a_0  G_3    \nonumber\\
& &
+ {\rm Tr}(\Pi \dot{A}_1)
+ {\rm Tr}({A}_0 {\cal D}_1 \Pi)
+ u^a \left [p_a 
- \frac{1}{4}
\left( V(\eta_A q^A)  
+ 8 \frac{{\rm Tr}(\Pi^2)}{ W(\eta_A q^A) }\right)
E_{a1} \right ]
\label{conpa}
\end{eqnarray}
The symplectic structure identifies the
canonical coordinates as $e_1^a$, $\omega_1$, $a_1$, $A_1$, and $q^a$,
while their respective canonical momenta are
${\eta_a \over 2 \pi \kappa}$,
${\eta_2 \over 2 \pi \kappa}$,
${\eta_3 \over 2 \pi \kappa}$,
$\Pi$, and $p_a$.   
The Hamiltonian is a superposition of
the gravitational gauge constraints
and Yang-Mills constraints;
the Lagrange multipliers
$e_0^a$, $\omega_0$, $a_0$ enforce the vanishing of the 
gravitational gauge generators,
$u^a$ enforces the vanishing of
\begin{eqnarray}
C_a \equiv 
\left [p_a - \frac{1}{4}
\left( V(\eta_A q^A)  
+ 8 \frac{{\rm Tr}(\Pi^2)}{ W(\eta_A q^A) }\right)
E_{a1} \right ]~, 
\label{ca}
\end{eqnarray}
and ${A}_0$ enforces the vanishing (Gauss constraint) of 
\begin{eqnarray}
{\cal G} \equiv {\cal D}_1 \Pi
~.
\label{ymgen}
\end{eqnarray}

\noindent
Using the Poisson brackets implied by the symplectic structure, one verifies
that the algebra of constraints closes; they are first-class.
The Yang-Mills generators follow the familiar Lie algebra
of the Yang-Mills group, which in components reads
\begin{eqnarray}
\left[ {\cal G}_a (x), {\cal G}_b (y) \right]_{PB} 
= f_{abc} {\cal G}_c (x) ~ \delta (x-y),
\end{eqnarray}
and the gravitational gauge generators  
follow the Lie algebra
\begin{eqnarray}
&&\left[ G_a (x), G_b (y) \right]_{PB} 
= \epsilon_{ab} G_3 (x) ~ \delta (x-y)
\label{eq13a} \\
&&\left[ G_a (x), G_2 (y) \right]_{PB} 
= \epsilon_{ab} G_b (x) ~ \delta (x-y) ~,
\label{eq13b}
\end{eqnarray}
where a common time argument has been suppressed.
Quantization consists of replacing Poisson brackets by
commutators.  We proceed following Ref.\cite{Can94}, i.e. 
we exploit the features of the  Schr\"odinger functional 
representation to postpone questions of 
the quantum nature of the constraint algebra
(\ref{eq13a})-(\ref{eq13b}).
We seek wave
functionals $\Psi$, in the Schr\"odinger representation,
that are solutions of the 
functional differential equations corresponding to
the requirement of vanishing constraints.
As done in Ref.\cite{Can94},
we do not well-order operators in the constraints at intermediate
steps of the calculation; the ordering is stipulated only at the end, 
when a constraint is taken to act on the wave functional.
We work in ``momentum'' space for the metric
and the Yang-Mills variables, {\it i.e.}
$\Psi$ depends on $\eta_a$, $\eta_2$, $\eta_3$, $\Pi$, and $q^a$ while
$e_1^a$, $\omega_1$, $a_1$, $A_1$, and $p_a$  
are realized as the functional derivatives
\begin{eqnarray}
&e_1^a \sim 2\pi \kappa i 
{\partial \over \partial \eta_a} ~,~~~
\omega_1 \sim 2\pi \kappa i 
{\partial \over \partial \eta_2} ~,~~~
a_1 2 \sim 2 \pi \kappa i 
{\partial \over \partial \eta_3} ~,~~~&
\nonumber \\
&A_1 \sim {1\over i} \, {\delta \over \delta \Pi}
~,~~~
p_a \sim {1\over i} \, {\delta \over \delta q^a} 
~,&
\label{momenta}
\end{eqnarray}
\noindent
Having clarified the objectives and methodology of our
investigation,
we can proceed in the investigation of the quantum mechanical theory.
We begin by observing that the $G_3 \! = \! 0$
constraint simply requires that the wave functional depends only on
the constant part of $\eta_3$, which we call $\lambda$
(in analogy with the classical theory, where it
corresponds to the cosmological constant).
We also notice that
\begin{equation}
{\eta^a \over \lambda} G_a - G_2
= {M' \over 4 \pi \kappa \lambda} 
+ \left( q^a + {\eta^a \over \lambda} \right)
\epsilon_a^{~~b} p_b
\label{a2project}
\end{equation}
where $M$ is the gauge-invariant combination\footnote{In the 
classical theory in absence of the Yang-Mills fields, 
$M$ is constant and corresponds to the ``black hole'' 
mass\cite{Can92,Can93,Can94}.}
\begin{equation}
M = \eta_a \eta^a - 2 \eta_2 \eta_3 ~.
\label{defm}
\end{equation}
Eq.(\ref{a2project}) implies that, 
as a result of the $G_a \! = \! 0$ and $G_2 \! = \! 0$ 
constraints, in the space of physical wave functionals 
the following operatorial relation holds
\begin{equation}
{M' \over 4 \pi \kappa \lambda} =
- \left( q^a + {\eta^a \over \lambda} \right)
\epsilon_a^{~~b} p_b ~.
\label{a2projectimplies}
\end{equation}
In light of the above observations, it is convenient to shift
some of the (functional) variables used to describe the wave
functional. The variables $q^a$ can be conveniently replaced
by the shifted variables
\begin{equation}
\rho^a \equiv q^a + \eta^a / \lambda ~,
\label{eq25}
\end{equation}
which respond only to Lorentz gauge transformations (they are
translation and $U(1)$ invariant).
$p_a$ can then be taken as
conjugate to $\rho^a$.
Moreover, $\eta_2$ can be shifted by $\eta_a \eta^a / (2 \lambda)$, 
so that $-2\lambda\eta_2$ is replaced by
the gauge invariant variable $M$. Correspondingly 
${1\over 2\pi \kappa}\omega_1$, the 
coordinate conjugate to $\eta_2$, is renamed
$2\lambda\Pi_M$, with $\Pi_M$ conjugate to $M$.

\noindent
With these redefinitions, one finds that
wave functionals satisfying the gravitational gauge constraints
take the form (also using the 
notations $\hat{\rho}^a \equiv \rho^a / \rho$ 
and $\rho \equiv \sqrt{\rho^a \rho_a}$)
\begin{equation}
\Psi = \delta({\eta_3} - \lambda) \, 
e^{i \Omega} \,
e^{i \widetilde{\Omega}} \,
\widetilde{\Psi} (M , \eta_3, \Pi, \rho^2)
\label{eq26}
\end{equation}
where $\Omega$ is the Kirillov-Kostant 1-form on the coadjoint orbit 
of the extended Poincar\'e group
\begin{equation}
\Omega = {1\over 
4 \pi \kappa \, \lambda} \int \epsilon^{ab} \eta_a d \eta_b
~,
\label{eq24}
\end{equation}
and 
\begin{eqnarray}
\widetilde{\Omega} = {1\over 4 \pi \kappa \lambda} \int d \hat{\rho}^a
\epsilon_{ab} \hat{\rho}^b M
~.
\label{eq31}
\end{eqnarray}
Once Eq.(\ref{a2project}) is taken into account,
the fact that (\ref{eq26}) solves the gravitational gauge
constraints can be easily checked in complete analogy with
corresponding analyses presented in Refs.\cite{Can92,Can93,Can94}.
The $\eta_a$-independence of $\widetilde{\Psi}$ can be traced back to 
the $G_a \! = \! 0$ constraint, whereas the vanishing of $G_2$ 
causes $\widetilde{\Psi}$ to depend on $\rho_a$ only through its
magnitude $\rho$.

\noindent
The structure of $\widetilde{\Psi}$ is further constrained by
the $C_a \! = \! 0$ and ${\cal G} \! = \! 0$ requirements,
which we have not yet imposed.
In particular, the ${\cal G} \! = \! 0$ constraint implies
$(\Pi^2)' \! = \! 0$, {\it i.e.} the wave functional depends only on
the constant part of $\Pi^2$, which we call $Q^2/16$ to be in agreement
with the conventions of sec. 3.

\noindent
Instead of imposing directly the constraint $C_a$, we equivalently
consider the combination
\begin{equation}
H_a\equiv C_a +\frac{1}{4\lambda} \hat V_Q(m) \epsilon^a_{~b} G^b=
p^a+\frac{1}{4}\hat V_Q(m)[\epsilon^a_{~b}\rho^{\prime b}+
\frac{2\pi\kappa}{\lambda}(p^a+2\lambda^2\rho^a \Pi_M)] \, ,
\end{equation}
where $\hat V_Q$ is the effective dilaton potential defined
in the previous section, and
we also introduced the notation
\begin{equation}
m \equiv \eta_A q^A = { \lambda^2 \rho^2 - M \over 2 \lambda} ~.
\label{etaaqa}
\end{equation}
To proceed it is convenient to decompose this constraint in its
radial and angular part, namely
\begin{equation}
H_a=\epsilon_{ab}\frac{\rho^b}{\rho^2} (\epsilon^{lm}\rho_l H_m)
+ \frac{\rho_a}{\rho^2} ( \rho^l H_l)
\end{equation}
with
\begin{eqnarray}
\label{giuda}
&&(\rho^l H_l)=\rho_a p^a +\frac{1}{4} \hat V_Q(m)[\epsilon_{ab}
\rho^{a}\rho^{'b}+\frac{2\pi\kappa}{\lambda}(\rho_a p^a+2\lambda^2 
\rho^2\Pi_M)].\\
\label{caproject}
&&(\epsilon^{lm}\rho_l H_m)=-\frac{1}{4\pi\kappa\lambda}[M^\prime-
\pi\kappa \hat V_Q(m) m^\prime]
\end{eqnarray}
In Eq.(\ref{caproject})  we have used Eq.(\ref{a2projectimplies}) to
eliminate the combination $\epsilon_{ab}\rho^a p^b$.

\noindent
Taking into account that $\rho_a p^a \widetilde \Omega=0$, 
the radial part (\ref{giuda})
leads to the following constraint
on $\widetilde\Psi$ 
\begin{equation}
\label{caproject11}
\rho_a p^a +\frac{\pi\kappa}{2\lambda}\hat V_Q(m)(\rho_a p^a+
2\lambda^2 \rho^2 \Pi_M).
\end{equation}
The structure of Eq.(\ref{caproject11})  suggests that
it might be convenient to shift the $M$ dependence in the remaining 
$\widetilde{\Psi}$ functional by the term  $-M(1+1/2\lambda)+\lambda
\rho^2/2$, which
is equivalent to a canonical transformation from $M$,
$\Pi_M$, and $p_a$ to $m$, 
$\Pi_m = - 2 \lambda \Pi_M$, and
$\Pi_a = p_a + 2 {\lambda^2} \rho_a \Pi_M$
($\rho^a$ is unaffected). In terms of the variables $m$ and $\rho^2$,
the constraint (\ref{caproject11}) on $\widetilde{\Psi}$ takes the form
\begin{equation}
\left [1-\frac{\pi\kappa}{2\lambda} \hat V_Q(m)\right]
\frac{\delta\widetilde \Psi}{\delta
\rho^2}+\lambda\frac{\delta\widetilde\Psi}{\delta m}=0.
\label{catilde}
\end{equation}
It is then easy to see that Eq.(\ref{catilde}) implies
that the wave functional only depends on the following
combination of the $m$ and $\rho^2$ variables:
\begin{equation}
\xi \equiv \lambda^2 \, \rho^2 
- 2  \, \lambda \, m -\pi \kappa J_Q(m)=(M-\pi \kappa J_Q(m))
~,
\label{catildeimplies}
\end{equation}
where $J_Q(x)$, as in the sec. 3, is a $x$-primitive of the 
function $\hat V_Q(x)$.

\noindent
Subsequently from (\ref{caproject}) it follows that only  the constant
part of $M \! - \pi \kappa J_Q(m)$ can appear nontrivially in the wave 
functional.

\noindent
Finally, we observe that the solutions of the ${\cal G} \! = \! 0$
constraint associated to the gauge field $A$ are of the form
\begin{equation}
e^{-i \Omega_\Pi} f_{GI}(\Pi)
~~~~\em{\rm with}~~ \Omega_\Pi 
= \int < K, g_{K \Pi}^{~} \, d g_{K \Pi}^{-1} >
~,
\label{solgauss}
\end{equation}
where we use the index ``$GI$'' to indicate that
$f_{GI}$ is a gauge-invariant functional of $\Pi$
({\it i.e.} depends only on the characters of the Yang-Mills group),
$<,>$ is the invariant inner product of the Lie algebra
of the Yang-Mills group,
$K$ is any fixed element of the Lie algebra,
and $g_{K \Pi}^{~}$ 
is a group element such 
that $\Pi \! = \! g_{K \Pi}^{-1} K g_{K \Pi}^{~}$.
For example, if the Yang-Mills group is $SU(2)$ 
(generators $\sigma_1$, $\sigma_2$, $\sigma_3$),
the reader can easily check that 
${\cal G}$ vanishes on the functional
\begin{equation}
e^{-i \Omega_\Pi^{SU(2)}} f_{GI}(\Pi)
~~~~\em{\rm with}~~ \Omega_\Pi^{SU(2)} 
= \int \Pi_2 {\Pi_1 \, d \Pi_3 - \Pi_3 \, d \Pi_1 \over \Pi_1^2 + \Pi_3^2}
~,
\label{solgaussu2}
\end{equation}
which corresponds to $K=\sigma_3$ and, accordingly,
\begin{eqnarray}
g_{K \Pi} = e^{i \theta_1 \sigma_1} e^{i \theta_2 \sigma_2}
~~~~\em{\rm with}~~ \theta_1 = - {1 \over 2} \arcsin(\Pi_2)
~,~~~\theta_2 = {1 \over 2} 
\arcsin \left({\Pi_1 \over \sqrt{\Pi_1^2+\Pi_3^2}} \right)
~.
\label{gsolu2}
\end{eqnarray}
The above analysis of the $C_a \! = \! 0$ and ${\cal G} \! = \! 0$ 
constraints
leads to the following final result for $\widetilde{\Psi}$
\begin{equation}
\widetilde{\Psi} =
\delta(\xi') \, \delta((\Pi^2)')
\, e^{-i \Omega_\Pi} 
\, \widetilde{\widetilde{\Psi}}_{GI}(\xi,\Pi)
~,
\label{solgaussWE}
\end{equation}
which together with Eq.(\ref{eq26}) gives the sought
physical wave funtional.

It is useful to consider some limiting cases of our result. 

\noindent
If $ V \! = \!  W \! = \! 0$,
the starting Lagrangian reduces
to the one of free gravity. It is not difficult to check that the correct
wave functional is recovered. In fact, $J_Q(m)\to 0$ in this limit, and only
the dependence on the constant part of $M$ survives in agreement 
with reference \cite{Can93,Can94}.

\noindent
If $ W \! = \! 0$ but
$ V \! \not = \! 0$, we
reproduce, by using the gauge formulation,
the results about the most general dilaton gravity obtained
in Ref.\cite{kun}.  Moreover it is interesting to notice  that for 
$V(x) \! = \! x^2$ the model 
is equivalent to the $R^2$ gravity with
the constraint of vanishing torsion. 

\noindent
If $ V \! = \! 0$ but $ W \! \not = \! 0$,
one can realize, by varying $W$,
a family of couplings involving metrics
that can be obtained by conformal rescaling
of the Cangemi-Jackiw gauge metric.
In particular, the case $W \! = \! \rm constant$ 
corresponds to Yang-Mills
fields minimally coupled to the gauge metric,
whereas $W(x) \! = \! {\rm exp} [ \int 
(d\phi/2) (d S^{\! _P}/d\phi)^{-1} ]$
corresponds to Yang-Mills fields minimally coupled to
the physical metric\cite{bs}.      

\pagebreak[3]
\setcounter{equation}{0}
\renewcommand{\theequation}{\arabic{section}.\arabic{equation}}
\section{Closing Remarks}
\nopagebreak
\medskip
\nopagebreak
\indent

Our analysis of string-inspired gravity coupled 
to Yang-Mills fields prompts several considerations.

Let us start by observing that the gauge theoretical
Cangemi-Jackiw formulation of string-inspired gravity
has led indeed to a very natural description 
of the coupling to Yang-Mills fields.
We reproduced and generalized several results
known in the geometrical formulation,
by showing that,
in the momentum representation,
the constraints could be straightforwardly enforced with the help
of a series of
Kirillov-Kostant phases.   

It should also be noticed that,
whereas in the realm of the geometrical formulation the 
class of theories here considered 
appears to be completely general,
in the gauge-theoretical formulation one can assume more complicated
structures for the potentials $V$ and $W$, in which they
depend on the additional two field variables $\xi$ and
$\eta_3$. [We remind the reader that $\xi$ and
$\eta_3$ are numbers in the geometrical formulation,
whereas they are independent scalar fields in the
gauge-theoretical formulation.]
Roughly speaking, a nontrivial dependence on $\xi$ allows  for 
both dynamical torsion and curvature, while a nontrivial 
dependence on $\eta_3$ ``turns on" the fields $a_\mu$ associated to 
the central extension. 
It would be interesting to investigate these more general scenarios.

Our results also indicate that 
Yang-Mills fields in two dimensions have no
dynamical degrees of freedom even when coupled to dilaton gravity.
As shown by Eq.(\ref{catildeimplies}), they only affect the geometry by 
modifying
the relation between the variable $M$
and the constant mode $\xi$
characterizing the wave functional
($\xi \! = \! M$ in pure gravity).
The solvability of the diffeomorphism constraints can be 
traced back to this topological nature of the
Yang-Mills fields, and the fact that, in such a context, 
these constraints
can be genuinely traded for the gauge-theoretical
constraints of the extended Poincar\'e group.
The difficulties encountered in Ref.\cite{Can94},
where the coupling of a scalar field to dilaton gravity
was investigated,
can be interpreted as a consequence 
of the disruption of the topological structure
caused by the dynamical degree of freedom
of the scalar field.

We also notice that in the context here considered,
unlike the case of dilaton gravity coupled to 
point particles\cite{bs}
no spurious divergences resulted from the use 
of the Poincar\'e variables.
This supports the interpretation\cite{bs} 
of the divergences encountered in the point-particle case
as a purely technical difficulty, originating from the fact
that the description of $N$ point particles requires the introduction
of $N$ sets of Poincar\'e variables with singularities associated
to the configurations with overlapping particle positions.
The description of
Yang-Mills fields coupled to dilaton gravity
requires the introduction of only one
set of Poincar\'e (field) variables.

We close by reemphasizing that our analysis should be
considered only as a first step toward the
challenging objective of a fully consistent quantization
of the gravity
Yang-Mills system.
By following the approach of Refs.\cite{Can92,Can93,Can94,kun},
we have postponed the issue of 
the quantum nature of the constraint algebra,
and the problem of defining a consistent
funtional measure in the space of {\it physical}
functionals that we identified.
This level of analysis 
has allowed us
to make a preliminary investigation of
the structure of the gravity Yang-Mills system,
leading in particular to the observation of
several differences between this system
and the previously investigated cases of
pure gravity\cite{Can92},
gravity coupled to point particles\cite{Can93,bs}, 
and gravity coupled to
scalar matter fields\cite{Can94}.
We hope that our results will be useful 
and will provide motivation for
future studies in which
the quantum nature of the problem be fully explored.

\bigskip
\begin{center}
{\bf ACKNOWLEDGEMENTS}
\end{center}
\ \indent
It is a pleasure to thank R. Jackiw for 
enlightening discussions.
\hfill
\eject

\end{document}